\documentclass[11pt,A4paper]{article}
\pdfoutput=1
\usepackage{jheppub}
\usepackage{amsmath}
\usepackage{amssymb}
\usepackage{hyperref}
\usepackage{graphicx}

\def \beq{\begin{equation}}
\def \eeq{\end{equation}}
\def \bea{\begin{eqnarray}}
\def \eea{\end{eqnarray}}

\def\bm#1{\mbox{\boldmath$#1$\unboldmath}}

\def \MString{M_{\rm string}}
\def \MSoft{M_{\rm soft}}
\def \MPlanck{M_{P}}
\def \compactification{string }

\preprint{OUTP-13-11P}

\title{Loop corrections to $\bm{\Delta N_{\rm eff}}$ in large volume models}

\author{Stephen Angus,}
\author{Joseph P. Conlon,}
\author{Ulrich Haisch,}
\author{Andrew J. Powell}

\affiliation{Rudolph Peierls Centre for Theoretical Physics, University of Oxford, \\
1 Keble Road, Oxford, OX1 3NP, United Kingdom}

\emailAdd{Stephen.Angus@physics.ox.ac.uk}
\emailAdd{j.conlon1@physics.ox.ac.uk}
\emailAdd{u.haisch1@physics.ox.ac.uk}
\emailAdd{Andrew.Powell2@physics.ox.ac.uk}

\abstract{In large volume models reheating is driven by the decays of the volume modulus to the visible sector, while the decays to its axion partners result in dark radiation.  In this article we discuss the impact  of  loop corrections on the only model-independent visible decay channel: the decay into Higgs pairs via a Giudice-Masiero term.  Including such radiative effects leads to a more precise determination of the relative fraction of dark radiation, since by contrast all loop corrections to the volume axion decay mode are Planck suppressed. Assuming an MSSM spectrum and that  the Giudice-Masiero coupling is fixed at the \compactification scale by a shift symmetry in the Higgs sector, we arrive at a prediction for the effective number of neutrinos.  The result turns out to be too large to be consistent with  data, highly disfavouring the minimal model.}

\begin{document}
\maketitle
\flushbottom

\section{Introduction}
\label{sec:intro}

The possible existence of dark radiation is interesting from both a theoretical and observational perspective. Observationally, dark radiation refers to an additional radiation density beyond that predicted in the $\Lambda$CDM model of standard Big Bang cosmology. It is conventionally parameterised by the effective excess number of neutrino species, $\Delta N_{\rm eff} \equiv N_{\rm eff} - 3.046$. Cosmic microwave background (CMB) experiments have developed increasing sensitivity to $\Delta N_{\rm eff}$ and further improvements are expected. When combined with the Hubble Space Telescope (HST) measurement of the Hubble constant $H_0$ \cite{Riess:2011yx}, the current results from the Wilkinson Microwave Anisotropy Probe (WMAP), the Atacama Cosmology Telescope (ACT), the South Pole Telescope (SPT) and Planck are $N_{\rm eff} = 3.84 \pm 0.40$ \cite{Hinshaw:2012aka}, $N_{\rm eff} =  3.71 \pm 0.35$ \cite{Hou:2012xq}, $N_{\rm eff} = 3.50 \pm 0.42$ \cite{Sievers:2013wk} and $N_{\rm eff} =  3.62 \pm 0.25$ \cite{Ade:2013lta}, respectively.\footnote{Without including direct measurements of $H_0$, the determinations using only CMB and baryon acoustic oscillations data are $N_{\rm eff} = 3.55 \pm 0.60$ (WMAP), $N_{\rm eff} = 2.87 \pm 0.60$ (ACT), $N_{\rm eff} = 3.50 \pm 0.47$ (SPT) and $N_{\rm eff} = 3.30 \pm 0.27$ (Planck).} These values are consistent with the independent bounds on $N_{\rm eff}$ from Big Bang Nucleosynthesis (BBN) alone, for which the latest value is  $N_{\rm eff} = 3.57 \pm 0.18$~\cite{13083240}. These measurements hint at an excess while clearly disfavouring new-physics scenarios that feature ${\cal O} (1)$ or larger contributions to~$N_{\rm eff}$.

Dark radiation is also interesting theoretically, as it is a simple and natural extension of $\Lambda$CDM. It is believed that after inflation the Universe was reheated from the decays of a scalar field. Dark radiation exists whenever this field has a non-zero branching ratio to light hidden-sector particles. An example of such a particle is the QCD axion (postulated to solve the strong CP problem), and other candidates include axion-like particles or hidden $U(1)$ gauge bosons. From this perspective, it is not a presence but an absence of dark radiation that would be a surprise -- dark radiation is only absent
if the reheating field has \emph{no} decay modes to hidden-sector particles.

Dark radiation also provides an arena to make contact between observations and models of Planck-scale physics. As matter redshifts more slowly than radiation, reheating is driven by the decays of the longest-living matter fields. (Any radiation produced by earlier decays will quickly redshift away and will have a subdominant contribution to the energy density of the universe.) These are naturally fields whose couplings are suppressed by inverse powers of the Planck mass $\MPlanck$. In string theory these are the moduli fields. The values of the coupling constants are all set by the vacuum expectation values (VEVs) of moduli, hence any potential acts as a source for moduli. In particular, the large energy densities during the inflationary epoch will displace the moduli VEVs from their final global minimum. After inflation the moduli oscillate about the global minimum as non-relativistic matter until they decay. The lightest moduli fields have the longest lifetimes, thus in theories with moduli it is expected that the lightest modulus field is responsible for reheating, almost independently of the details of either the inflationary model or the rest of the particle spectrum.


The significance of this is that entropy production as well as reheating will dominantly come from the last field to decay, independent of how many other quickly-decaying fields may exist after inflation. The lightest modulus field is the most long-lived field with only gravitational couplings, hence it is the last field to decay. It is thus this field alone which sets the reheating temperature. Indeed, this is the essence of the moduli problem \cite{Coughlan, BanksNelson, Quevedo}: models of low-energy supersymmetry (SUSY) within string theory generally have moduli not much heavier than the TeV scale, but unless the moduli masses are above roughly $30 \, \hbox{TeV}$ then they decay after BBN with an unacceptably low reheating temperature $T_{\rm reheat}$.

One string model of dark radiation was studied in \cite{Cicoli:2012aq,Higaki:2012ar} within a sequestered form of the large volume scenario (LVS) \cite{Balasubramanian:2005zx,Conlon:2005ki}  (see also \cite{Higaki:2013lra} for more general considerations). The large volume scenario is tractable to analyse as it has a unique lightest modulus,~{i.e.}~the volume modulus $\Phi$, which is parametrically lighter than any other modulus. As argued above, the presence of a single lightest Planck-coupled modulus then implies that within these models reheating should be driven by decays of the volume modulus,
independently of the details of the high-scale inflationary model.

This is attractive as, being the volume modulus, the majority of its couplings are calculable in a model-independent fashion. In fact, there are two important couplings. The first is to the volume axion $a_b$, which is a hidden-sector state and the corresponding decay channel $\Phi \to a_b \hspace{0.25mm} a_b$ hence gives rise to dark radiation. The second coupling is to the bilinear $H_u H_d$ of Higgs fields. This interaction leads to the only competitive visible-sector decay mode, $\Phi \to H_u H_d$, and induces the reheating of the Universe. The corresponding coupling $Z$ is an undetermined constant with a natural value of ${\cal O}(1)$ at the \compactification scale $\MString$. However, if the Higgs sector has an exact shift symmetry (see the recent works~\cite{Hebecker:2012qp,Hebecker:2013lha} for explicit string theory constructions of such a symmetry), then $Z$ is fixed to 1 at $\MString$. The case of a shift-symmetric Higgs sector with pure MSSM matter content is then completely defined and predictive. We will refer to this specific LVS  as minimal LVS (MLVS).

In the MLVS the ratio of branching ratios of visible-sector and hidden-sector decays is simply given by  \cite{Cicoli:2012aq,Higaki:2012ar}
\beq \label{eq:ratioBRs}
\kappa \equiv \frac{{\rm Br} ({\rm hidden})}{{\rm Br} ({\rm visible})} = \frac{{\rm Br} (\Phi \to a_b \hspace{0.25mm} a_b)}{{\rm Br} (\Phi \to H_u H_d)} = \frac{1}{2 Z^2} \,,
\eeq
where the coupling $Z$ is understood to be normalised at the mass $m_\Phi$ of the volume modulus. In terms of (\ref{eq:ratioBRs}) the  effective excess number of neutrinos is given by  $\Delta N_{\rm eff} \simeq 3.3 \hspace{0.5mm}  \kappa$ \cite{Cicoli:2012aq,Higaki:2012ar}. At tree level one has $Z(m_\Phi) = Z(\MString)=1$, which implies $\kappa = 1/2$ and $\Delta N_{\rm eff} \simeq 1.7$.  On the other hand, the measured values of $N_{\rm eff}$ require $\Delta N_{\rm eff} \lesssim 1.1$, which translates into $\kappa \lesssim 1/3$.  The MLVS tree-level prediction for $\Delta N_{\rm eff}$  is hence in conflict with observation.

However, even if the Higgs sector is exactly shift symmetric at the compactification scale, this symmetry is broken by the gauge and Yukawa couplings. In consequence,  the coupling $Z$  will receive logarithmically-enhanced corrections of the form  $1/(4 \pi) \ln \left (\MString/m_\Phi \right )$ from MSSM loop diagrams. In view of the large hierarchy $\MString \gg m_\Phi$, the resulting terms can be of ${\cal O} (1)$ and have to be resummed using renormalisation group (RG) techniques. An immediate question then arises as to whether the induced radiative corrections are large enough to make the MLVS compatible with the measurements of $N_{\rm eff}$, which call for $Z(m_\Phi) \gtrsim 1.2$. The purpose of our paper is to answer this question.

Let us make some general comments on the cosmology implicit in the model under consideration. Since reheating arises from modulus decays, the reheating temperature is $T_{\rm reheat} \sim 1 \, \hbox{GeV}$. As emphasised above, low reheating temperatures are common to all models of SUSY breaking once they are embedded into string theory and moduli are included into the spectrum. As the temperature is lower than the  decoupling temperature of  weakly interacting massive particles (WIMPs), the conventional freeze-out calculation of MSSM neutralino dark matter (DM) become invalid. WIMP DM can however still be produced, for instance via non-thermal production in modulus decays. Another natural DM candidate are axions, for which late modulus decays dilute the axion DM abundance and allow higher values of the axion decay constant $f_a$ than in a conventional cosmology. A potentially more serious problem is the compatibility of baryogenesis with low-scale reheating. While this is beyond the scope of this paper, we note that Affleck-Dine baryogenesis has been argued to give acceptable baryon asymmetries even for low reheating temperatures \cite{hepph9503303}.

Our work is organised as follows. After reviewing some of the basic ingredients of the LVS, we calculate in  Section \ref{sec:analytic} the anomalous dimension of the coupling $Z$. Our analytic computation is complemented in Section \ref{sec:numerics} by a numerical RG analysis of the $\Phi \to H_u H_d$ decay mode, including one-loop and two-loop effects. We conclude in Section~\ref{sec:conclusions}.

\section{Analytic results}
\label{sec:analytic}

Before turning to the calculation of the anomalous dimension of the coupling  $Z$ that determines the relative amount of dark radiation via (\ref{eq:ratioBRs}), let us emphasise the main assumptions that enter our analysis. First, the general spectrum of moduli masses is set by the LVS. Second, the volume modulus is displaced from its eventual minimum during inflation. Given this, the volume modulus will come to dominate the energy density of the Universe and thereby drive reheating. Third, the spectrum of SUSY breaking soft terms has the sequestered form given in \cite{Blumenhagen:2009gk}. If sequestering were not realised, the TeV soft masses would require the volume modulus to be  light, which would lead to the cosmological moduli problem.  We now review some important features of the LVS relevant to our work.

\subsection{Mass hierarchies}
\label{sec:scales}

In the LVS the volume ${\cal V}$ of the Calabi-Yau manifold is stabilised at exponentially large values~\cite{Balasubramanian:2005zx}. This stabilisation mechanism creates a naturally small expansion parameter,~{i.e.}~the inverse volume, and leads to a distinctive hierarchy of scales, given in the sequestered LVS by~\cite{Blumenhagen:2009gk}
\beq \label{eq:scales}
\MString \sim \frac{\MPlanck}{{\cal V}^{1/2}} \,, \qquad
m_{\Phi} \sim \frac{\MPlanck}{{\cal V}^{3/2}} \,, \qquad
\MSoft \sim \frac{\MPlanck}{{\cal V}^{2}} \,, \qquad
m_{a_b} \lesssim \MPlanck \hspace{0.5mm} e^{-2 \pi {\cal V}^{2/3}} \sim 0 \,.
\eeq
Here $\MPlanck = 2.4 \cdot 10^{18} \, {\rm GeV}$ is the reduced Planck mass, while $\MString$, $m_\Phi$, $\MSoft$ and $m_{a_b}$ denote the \compactification scale, the mass of the volume modulus $\Phi$, the scale of the SUSY breaking soft masses and the mass of the volume axion $a_b$, respectively.  In what follows, we will assume that the level of volume sequestering is the same for both  scalar and gaugino masses, so that  $\MSoft \sim m_0 \sim m_{1/2}$. To solve the gauge hierarchy problem,~{i.e.}~$\MSoft \sim 1 \, {\rm TeV}$, one needs ${\cal V} \sim 5 \cdot 10^7$, resulting in $\MString \sim 3 \cdot 10^{14} \, {\rm GeV}$ and $m_\Phi \sim 7 \cdot 10^{6} \, {\rm GeV}$.

\subsection{Volume modulus interactions}
\label{sec:interactions}

The interaction terms in the LVS Lagrangian that give rise to the leading decay modes of the volume modulus,~{i.e.}~$\Phi \to a_b \hspace{0.25mm} a_b$ and $\Phi \to H_u H_d$, are
\beq \label{eq:L}
\begin{split}
{\cal L} &   \supset \frac{2}{\sqrt{6} \MPlanck} \, (\partial_\mu a_b)^2 \, \Phi +  \frac{1}{\sqrt{6} \MPlanck} \, \Big [ Z  \hspace{0.25mm} H_u H_d  \hspace{0.5mm} \Box \hspace{0.25mm}  \Phi + {\rm h.c.} \Big  ] \,,
\end{split}
\eeq
where $H_u H_d = \epsilon_{ij} H_u^i H_d^j = H_u^+ H_d^- - H_u^0 H_d^0$ represents the usual  contraction between $SU(2)_L$ doublets and all fields are understood to be canonically normalised. The second  term in  the above formula arises from a Guidice-Masiero term  \cite{Giudice:1988yz}  in the K\"ahler potential. While the presence and the form of the $\Phi \hspace{0.25mm} a_b  \hspace{0.25mm} a_b$ and $\Phi \hspace{0.25mm}  H_u H_d$  couplings in (\ref{eq:L}) are robust predictions of the LVS~\cite{Cicoli:2012aq,Higaki:2012ar}, in generic models further contributions  to the volume modulus decays into the hidden and visible sector may arise. Contributions of the former type can come~{e.g.}~from local closed string axions, but since such decays represent dark radiation they will always enhance the ratio $\kappa$ introduced in (\ref{eq:ratioBRs}),  worsening the tension between theory and experiment. Thus we will not consider additional hidden-sector contributions beyond $\Phi \to a_b \hspace{0.25mm} a_b$ here. Decays of the volume modulus into gauge bosons, MSSM scalars or both SM and MSSM fermions are expected to be either loop suppressed or mass suppressed (since $m_t < m_{0,1/2} \ll m_\Phi$), rendering the channel $\Phi \to H_u H_d$ the dominant visible MSSM decay mode.

The  partial decay rates of the volume modulus induced by  (\ref{eq:L}) read \cite{Cicoli:2012aq,Higaki:2012ar}
\beq \label{eq:Gammas}
\Gamma (\Phi \to a_b \hspace{0.25mm} a_b) = \frac{1}{48 \pi} \frac{m_\Phi^3}{\MPlanck^2} \,, \qquad
\Gamma (\Phi \to H_u \hspace{0.25mm} H_d) = \frac{2 Z^2}{48 \pi} \frac{m_\Phi^3}{\MPlanck^2} \,,
\eeq
where again $Z = Z(m_\Phi)$. From these results the expression (\ref{eq:ratioBRs}) follows readily.

\subsection{Running of volume modulus Higgs coupling}
\label{sec:rge}

In contrast to the $\Phi \hspace{0.25mm} a_b  \hspace{0.25mm} a_b$ coupling, which receives only Planck-suppressed corrections, the $\Phi \hspace{0.25mm}  H_u H_d$ coupling is modified by the virtual exchange of MSSM particles. As a result the interaction strength $Z$ entering (\ref{eq:L}) will evolve logarithmically from $\MString$ to $m_{\Phi}$, where the volume modulus decays. The scale dependence of $Z$ is determined by the following RG equation
\beq \label{eq:RGEZ}
\frac{d}{d t} Z = \gamma_Z Z \,,
\eeq
where $t \equiv \ln \left (Q/Q_0 \right )$ with $Q$ denoting the renormalisation scale and $Q_0$ a reference scale, and $\gamma_Z$ is the corresponding anomalous dimension.

\begin{figure}[!t]
\centering
\includegraphics[height=0.45\textwidth]{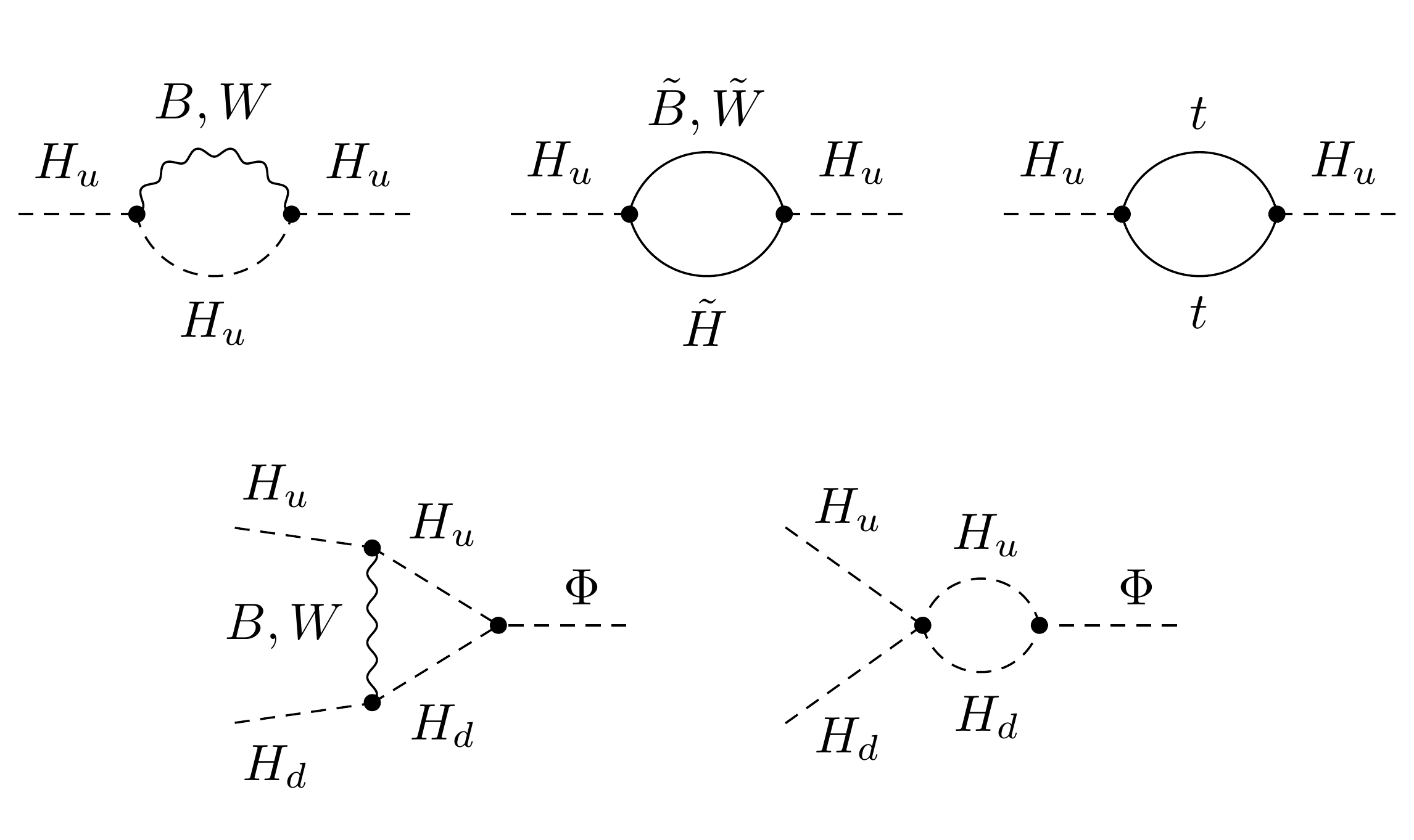}
\caption{The $H_u H_u$ self-energy diagrams (upper row) and $\Phi  \hspace{0.25mm}   H_u H_d$ vertex diagrams (lower row) that contribute to the one-loop anomalous dimension $\gamma_Z^{(1)}$. The $H_d H_d$ self-energy diagrams needed to determine the wave-function renormalisation factor of $H_d$ are not shown.}
\label{fig:diagrams}
\end{figure}

Since the $\Phi \hspace{0.25mm}  H_u H_d$ coupling arises from the K\"ahler potential and the volume modulus field itself does not renormalise, the anomalous dimension $\gamma_Z$ can be written
in terms of Higgs wave-function renormalisations  (as a consequence of the supersymmetric non-renormalisation theorem \cite{Grisaru:1979wc}). One obtains
\beq \label{eq:gammaZ}
\gamma_Z = \gamma_{H_u} + \gamma_{H_d} \,,
\eeq
where $\gamma_{H_u}$ and $\gamma_{H_d}$ are the anomalous dimensions of the Higgs superfields. To verify the correctness of (\ref{eq:gammaZ}), we have calculated the one-loop correction $\gamma_Z^{(1)}$ to the anomalous dimension $\gamma_Z$ explicitly.  The corresponding Feynman diagrams are depicted in Figure~\ref{fig:diagrams}. We performed the calculation of the self-energy and vertex diagrams using dimensional regularisation with modified minimal subtraction (i.e.~the $\overline{\rm DR}$ scheme). The contributions to the scalar fields $H_{u,d}$ were computed in Wess-Zumino gauge, retaining an arbitrary $R_\xi$ gauge for the vector fields.\footnote{Note that in Wess-Zumino gauge SUSY is broken, which explains the need to compute vertex diagrams.} While both classes of diagrams are individually gauge dependent, the $\xi$ dependence cancels in the sum of contributions. Our results for the individual diagrams agree with those given in \cite{Machacek:1983tz, Machacek:1983fi, Capper:1984zy}.  Keeping only the third-family Yukawa couplings $y_{t,b,\tau}$, we obtain
\beq \label{eq:gammaZ1}
\gamma_Z^{(1)} =  \frac{1}{(4\pi)^2}  \left [ -\frac{3 g_1^2}{5}  - 3 g_2^2 + 3  \hspace{0.25mm}  | y_t |^2 +  3  \hspace{0.25mm}  |y_b|^2 + |y_\tau|^2 \right ] \,,
\eeq
which equals the sum $\gamma_{H_u}^{(1)} + \gamma_{H_d}^{(1)}$ of one-loop superfield anomalous dimensions as given~{e.g.}~in the review~\cite{Martin:1997ns}. Here the couplings $g_1$ and $g_2$ are given in terms of the conventional $U(1)_Y$ and $SU(2)_L$ SM gauge couplings by $g_1 = \sqrt{5/3} \hspace{0.5mm} g^\prime$ and $g_2 = g$, respectively.

Employing the one-loop anomalous dimension  (\ref{eq:gammaZ1}) to solve the RG equation (\ref{eq:RGEZ}), we find to  leading logarithmic accuracy
\beq \label{eq:K}
K \equiv \frac{Z(m_\Phi)}{Z (\MString)} \simeq 1 -\gamma_Z^{(1)}  \, \ln \left ( \frac{\MString}{m_\Phi} \right )  \simeq 1 - \frac{18}{(4 \pi)^2 } \left (  -1.7 +  \frac{1.5}{\sin^2 \beta} +  \frac{1.6 \cdot 10^{-4}}{\cos^2 \beta}  \right )   \,.
\eeq
To arrive at the numerical expression we have employed $g_1 (\MString) \simeq 0.65$, $g_2 (\MString) \simeq 0.69$, $y_t (\MString) \simeq 0.70/\sin \beta$,  $y_b (\MString) \simeq 6.0 \cdot 10^{-3}/\cos \beta$ and $y_\tau (\MString) \simeq 7.2 \cdot 10^{-3}/\cos \beta$, corresponding to $\MString = 3 \cdot 10^{14} \, {\rm GeV}$ and $m_\Phi =  7 \cdot 10^{6} \, {\rm GeV}$.\footnote{These values for the couplings have been obtained using the numerical procedure outlined in Section~\ref{sec:solution}.} In the final result  in~(\ref{eq:K}), we have shown the contributions arising from  the terms $g_{1,2}^2$, $|y_t|^2$ and $|y_{b,\tau}|^2$ separately. The  different overall signs multiplying the contributions from the gauge and the Yukawa couplings imply that the individual terms in (\ref{eq:gammaZ1}) tend to cancel. In fact,  the numerical expression for $\gamma_Z^{(1)}$ used in~(\ref{eq:K}) is less than 0 for $3 \lesssim \tan \beta \lesssim 35$ and vice versa. We hence expect to find that loop corrections suppress (enhance) the partial decay rate $\Gamma (\Phi \to H_u H_d)$ for small and large (moderate) ratios of the Higgs vacuum expectation values, $\tan \beta$.  In order to obtain a reliable prediction for $K$, however, the large logarithm appearing in~(\ref{eq:K}) has  to be resummed by solving (\ref{eq:RGEZ})  together with the RG equations describing the scale dependence of the gauge and Yukawa couplings.

\section{Numerical results}
\label{sec:numerics}

After presenting the analytic result for the anomalous dimension of the Guidice-Masiero coupling, we now turn to the numerical RG analysis of the $\Phi \to H_u H_d$ decay mode.  Our methodology is  detailed in the following.

\subsection{Solution of RG equations}
\label{sec:solution}

The system of differential equations describing the renormalisation scale dependence of the coupling strength $Z$ as well as those of the gauge and Yukawa couplings is solved iteratively with the help of {\sc SoftSusy~3.3.7}~\cite{Allanach:2001kg}. The calculation is performed including all relevant one-loop and two-loop effects.\footnote{Two-loop effects have however a very minor impact on the analysis, and therefore we only reported the result of the one-loop anomalous dimension $\gamma_Z^{(1)}$ in (\ref{eq:gammaZ1}).} The fine structure constant $\alpha (m_Z) = 1/127.973$, the Fermi constant $G_F = 1.16637 \cdot 10^{-5} \, {\rm GeV}^{-2}$, the strong coupling $\alpha_s (m_Z)$, the pole mass $m_t$ of the top quark, the bottom mass $m_b = 4.2 \, {\rm GeV}$ and the tau mass $m_\tau = 1.777 \, {\rm GeV}$ serve as SM inputs and constraints in the RG evolution. The low-energy boundary conditions are applied at the $Z$-boson mass $m_Z = 91.1875 \, {\rm GeV}$. At the \compactification scale $\MString$ we impose minimal supergravity (MSUGRA)  boundary conditions, which  just leaves five free SUSY parameters:  common scalar and gaugino masses, $m_0$ and $m_{1/2}$, universal trilinear terms $A_0$, the bilinear soft SUSY breaking term $B$ and the SUSY $\mu$ parameter.  Following common practice, we use the one-loop corrected electroweak symmetry breaking (EWSB) conditions (see~{e.g.}~\cite{Pierce:1996zz}) to trade $B$ and the magnitude $|\mu|$ in favour of $\tan \beta$ and the sign of $\mu$.  Notice that the assumed scaling of $m_0 \sim m_{1/2} \sim  \MPlanck/{\cal V}^2$ naturally requires $B \sim \MPlanck^2/{\cal V}^4$ and $\mu \sim \MPlanck/{\cal V}^2$ to achieve EWSB. We assume that these scalings are realised by an appropriate volume sequestering, and furthermore take $A_0 \sim \MPlanck/{\cal V}^2$. The SUSY scale is determined by the geometric mean $\MSoft \equiv \sqrt{m_{{\tilde t}_{1}} m_{{\tilde t}_{2}}}$ of the masses $m_{{\tilde t}_{1,2}}$ of the  stop mass eigenstates. Finally, the mass of the volume modulus is obtained from $m_\Phi = \MString^3/\MPlanck^2$.

\subsection{SM and MSUGRA parameter dependencies}
\label{sec:parameter}

Before studying the dependencies of  (\ref{eq:K}) on the MSUGRA parameters we consider the impact  of the parametric SM errors. The dominant sources of SM uncertainties arise from the top mass and the strong coupling constant. This is to be expected because~(\ref{eq:gammaZ1}) is quadratic in the top Yukawa coupling  and the RG evolution of  $y_t$ depends sensitively on the low-energy initial conditions for  $m_t$ and $\alpha_s$. The more critical ingredient is the top mass for which the latest Tevatron measurements  find $m_t = (173.2 \pm 0.9) \, {\rm GeV}$~\cite{Aaltonen:2012ra}. However,  the exact meaning of the mass parameter measured by CDF and D0 via a kinematical reconstruction of the top decay products and comparison to Monte Carlo simulations is unclear and so is its connection to $y_t$. A theoretically well-defined determination of $m_t$ can, on the other hand, be obtained from the total cross section for top-quark pair production. While such extractions  (see~{e.g.}~\cite{Beneke:2011mq})  give values for $m_t$ that are compatible with the mass determinations from direct reconstruction, the achieved accuracy is notably worse, with an uncertainty of around $\pm 5 \, {\rm GeV}$. The world average of the strong coupling  evaluated at the  $Z$-boson mass is $\alpha_s (m_Z) = 0.1184 \pm 0.0007$~\cite{Beringer:1900zz}. This value of $\alpha_s$ is obtained from a large set of measurements with significant spreads between them. To account for this fact we will also give results employing the $3 \sigma$ error $\pm 0.0021$ of the $\alpha_s$ world average.

\begin{figure}[!t]
\centering
\includegraphics[height=0.45\textwidth]{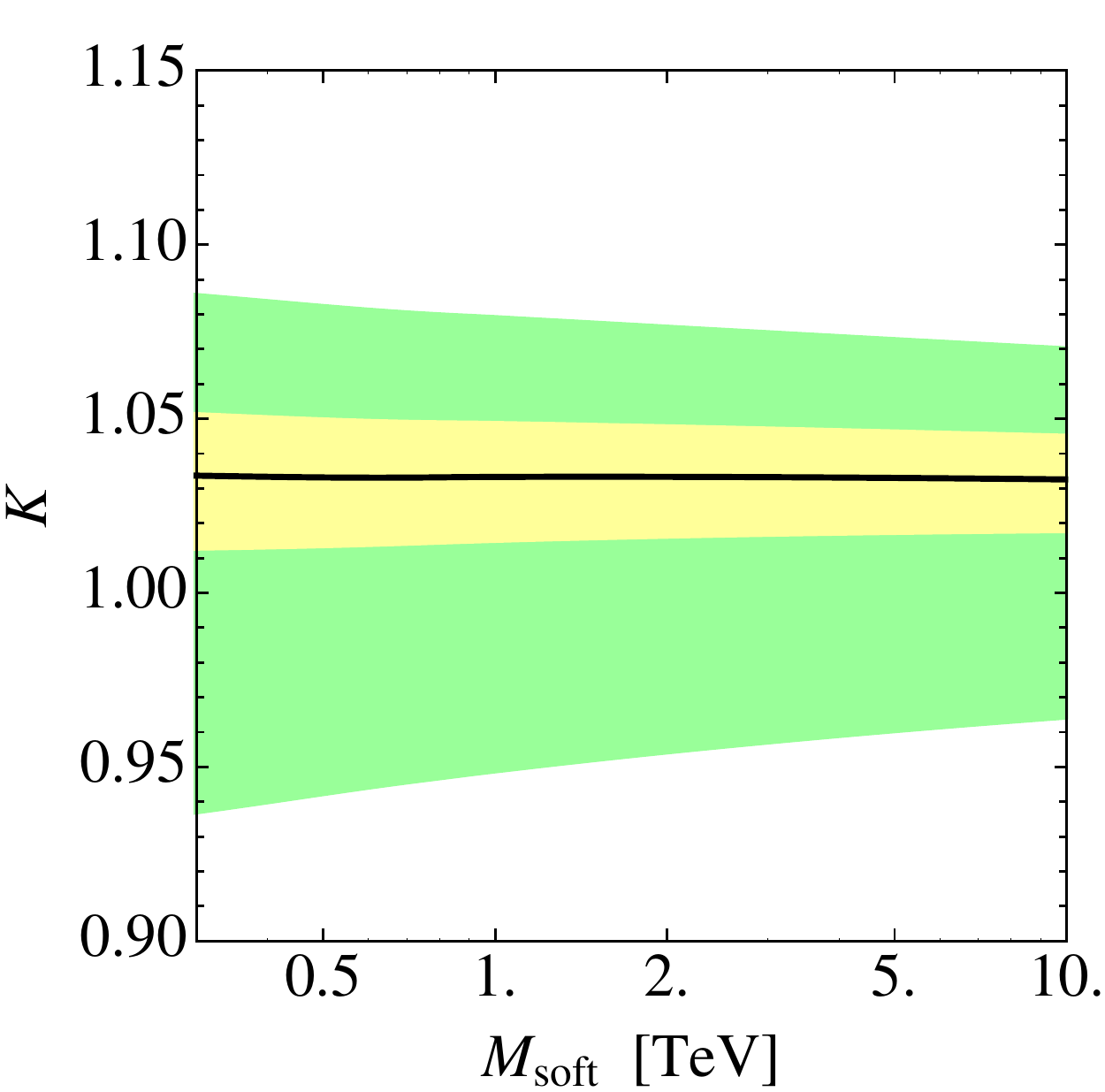} \qquad
\includegraphics[height=0.45\textwidth]{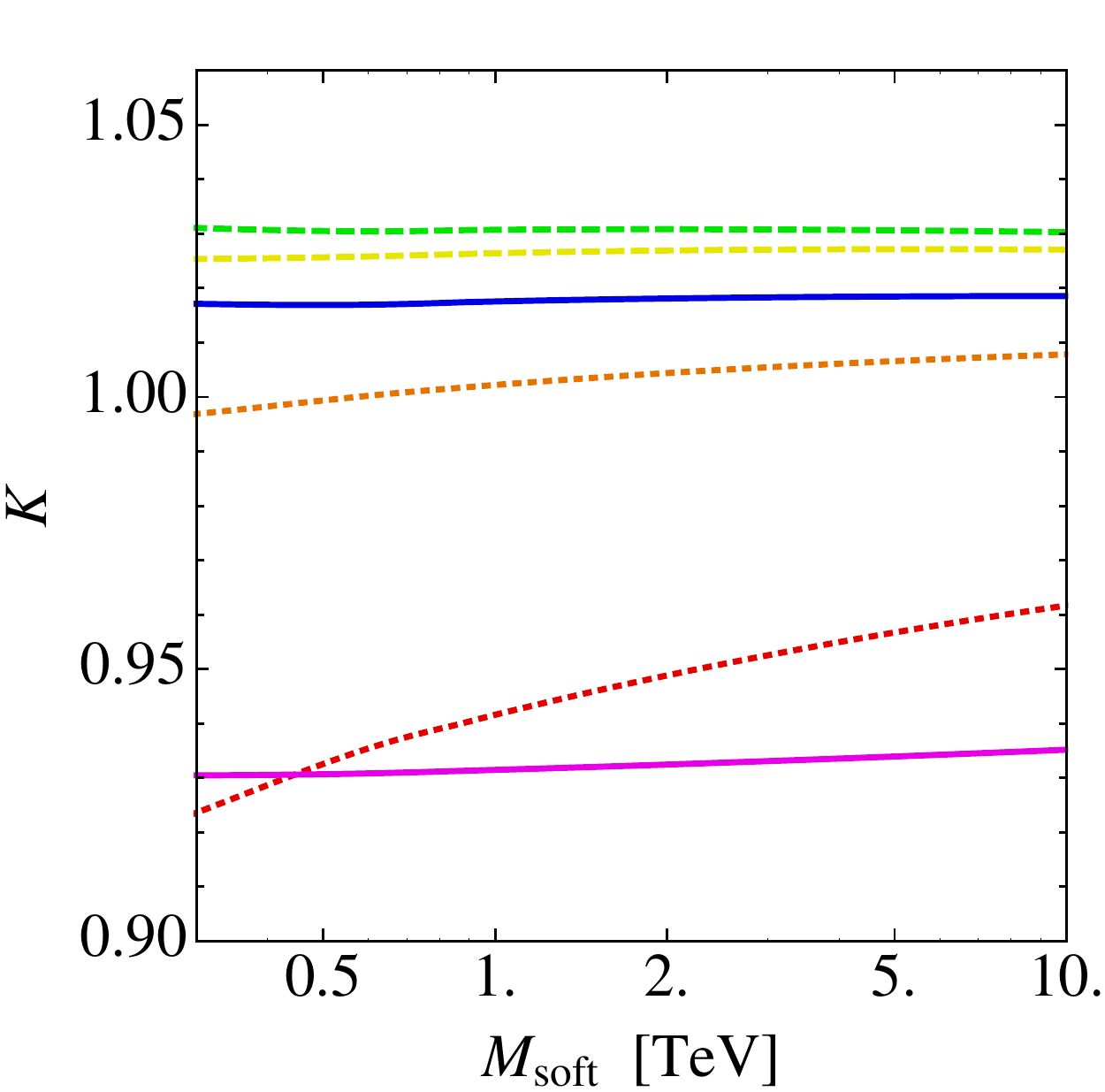}
\caption{Left: Predictions for $K$ for fixed MSUGRA input. The solid black line indicates the result obtained for the central choice of SM inputs while the coloured bands reflect the uncertainties associated with the errors in the top mass and the strong coupling constant. Right:  Predictions for $K$ for different values of $\tan \beta$. See text for further explanations.}
\label{fig:parameter}
\end{figure}

Our predictions for $K$ as a function of $\MSoft$ are shown in Figure~\ref{fig:parameter}. The results displayed in the left panel correspond to $m_0 = m_{1/2} = A_0$, $\tan \beta = 10$ and ${\rm sign} \hspace{0.5mm} \mu = +1$. Almost identical predictions are obtained for different choices of $A_0$ and ${\rm sign} \hspace{0.5mm} \mu = -1$. The solid black curve corresponds to $m_t = 173.2 \, {\rm GeV}$ and $\alpha_s (m_Z) = 0.1184$, while the yellow (green) band has been obtained by varying $m_t$ and $\alpha_s (m_Z)$  by $\pm 0.9 \, {\rm GeV}$ and $\pm 0.0007$ ($\pm 5 \, {\rm GeV}$ and $\pm 0.0021$) around their central values. We see that the ratio (\ref{eq:K}) is largely independent of the SUSY scale $\MSoft$, but that the exact value of $K$ depends to some extent on the low-energy input  $m_t$ and $\alpha_s (m_Z)$. Numerically, we find that the variations of $\pm 0.9 \, {\rm GeV}$ and $\pm 0.0007$ ($\pm 5 \, {\rm GeV}$ and $\pm 0.0021$) lead to shifts in $K$ of less than $\pm 2\%$ ($^{+5\%}_{-10\%}$) relative to the central values. The largest  value of (\ref{eq:K}) is thereby attained for the smallest value of $m_t$ and the largest value of $\alpha_s (m_Z)$, and vice versa.

We now analyse the dependence of $K$ on the choice of $\tan \beta$. Our numerical results are shown  in the right panel of Figure~\ref{fig:parameter}. All curves have been obtained for  $m_0 = m_{1/2} = A_0$, ${\rm sign} \hspace{0.5mm} \mu = +1$, $m_t = 173.2 \, {\rm GeV}$ and $\alpha_s (m_Z) = 0.1184$. The dotted red, dotted orange, dashed yellow, dashed green, solid blue and solid magenta  lines correspond to $\tan \beta = 2, 3, 5, 15, 25$ and $50$, respectively.  As anticipated, we find  that  for $\tan \beta \lesssim 3$ the predictions for the ratio~(\ref{eq:K}) are below~1, while for moderate values of $\tan \beta$ one obtains ratios   above 1. In fact, the values of $K$ saturate for $\tan \beta \simeq 10$ and increasing $\tan \beta$ further leads to a suppression of the ratio $\big ($a feature that is also reproduced by the simple formula (\ref{eq:K})$\big)$. For large $\tan \beta$ values, the ratio $K$ then ends up below 1. We  see furthermore that varying $\tan \beta$ in the range $[2,50]$ shifts $K$ by only $^{+3\%}_{-7\%}$ away from 1. The dependencies on the other MSUGRA parameters are even less pronounced than that on $\tan \beta$.

\subsection{Predictions for the effective excess number of neutrinos}
\label{sec:higgs}

It is well-known that the mass $m_h$ of the Higgs boson puts stringent constraints on the MSUGRA parameter space. This is particularly true after the discovery of a relatively heavy Higgs-like state with a mass of around $126 \, {\rm GeV}$ by ATLAS \cite{Aad:2012tfa} and CMS  \cite{Chatrchyan:2012ufa}. This feature can be easily understood by recalling the classic MSSM result for $m_h$ \cite{Casas:1994us,Dabelstein:1994hb} that includes the dominant one-loop contributions arising from an incomplete cancellation of top-quark loops and top-squark loops. It reads
\beq \label{eq:mh2}
m_h^2 \simeq m_Z^2 \hspace{0.25mm} \cos^2 \left ( 2\beta \right ) + \frac{3  G_F  \hspace{0.25mm} m_t^4}{\sqrt{2} \pi^2} \left [ \ln \left ( \frac{\MSoft^2}{m_t^2} \right )+ \frac{X_t^2}{\MSoft^2} \left ( 1 - \frac{X_t^2}{12 \MSoft^2} \right )  \right ] \,,
\eeq
where $X_t \equiv A_t - \mu \cot \beta$ denotes the stop-mixing parameter and $A_t$ is the trilinear stop-Higgs coupling.  The first term in (\ref{eq:mh2}) encodes the tree-level contribution to the squared mass of the Higgs and is maximised for $\tan \beta \to  \infty$, while the second term approximates the one-loop corrections and is maximised for $X_t  = \pm \sqrt{6} \hspace{0.25mm} \MSoft$ (known as maximal mixing).

\begin{figure}[!t]
\centering
\includegraphics[height=0.45\textwidth]{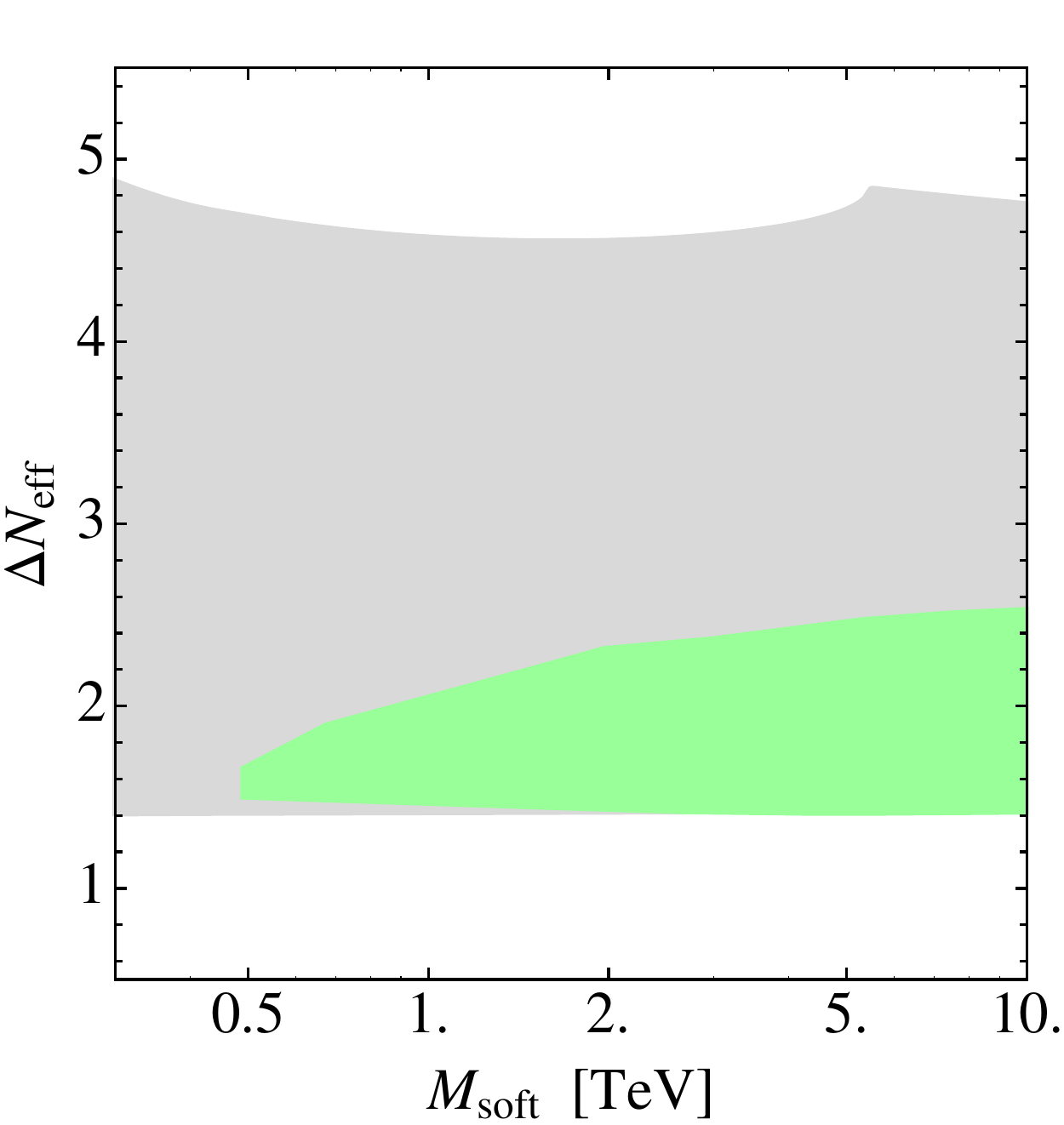}
\caption{Predictions for the effective excess number of neutrino species in the MLVS framework. The coloured wedge-shaped region indicates the possible values of $\Delta N_{\rm eff}$ consistent with the LHC measurements of a Higgs-like state near $126 \, {\rm GeV}$. For comparison the accessible parameter space without imposing the Higgs constraint is underlaid in grey. For further details see text.}
\label{fig:DNeff}
\end{figure}

We assess the impact of the LHC measurements of the Higgs mass on the predictions for $\Delta N_{\rm eff}$ by performing a global scan in the MSUGRA parameter space. Only points that lead to $m_h \in [123, 129] \, {\rm GeV}$ are retained, which is the range allowed by the ATLAS and CMS data~\cite{Aad:2012tfa,Chatrchyan:2012ufa} if one accounts for the theoretical uncertainties in the MSSM calculation of the Higgs mass (see~{e.g.}~\cite{Allanach:2004rh}). We generate a large sample of points allowing the MSUGRA parameters to take random values within $m_{0,1/2} \in  [0.1, 10] \, {\rm TeV}$, $A_0 \in [-30, 30] \, {\rm TeV}$ and $\tan \beta \in [1,60]$, permitting $\mu$ to be of either sign. In order to incorporate SM uncertainties we let the top mass and the strong coupling constant float within $m_t = (173.2 \pm 5)\, {\rm GeV}$ and $\alpha_s (m_Z) = 0.1184 \pm 0.0021$, respectively.  The allowed range of $\Delta N_{\rm eff}$ as a function of $\MSoft$ is then found from the minimal and maximal  $\Delta N_{\rm eff}$ values that are consistent with \cite{Cicoli:2012aq},
\beq \label{eq:Neff2bound}
3.12 \hspace{0.5mm} \kappa \leq \Delta N_{\rm eff} \leq 3.48  \hspace{0.5mm} \kappa \,,
\eeq
where $\kappa$, as defined in (\ref{eq:ratioBRs}), is calculated for each point. Notice that the two-sided bound~(\ref{eq:Neff2bound}) takes into account the uncertainty associated with the value of the reheating temperature. Since we effectively scan over all individual sources of uncertainties, the derived limits on $\Delta N_{\rm eff}$ should be considered very conservative.

Our results of the  MSUGRA scan are shown in Figure \ref{fig:DNeff}. The accessible parameter before (after) imposing the Higgs-mass constraint is indicated by the grey (coloured) region.  We see that in the MLVS the values for $\Delta N_{\rm eff}$ compatible with the $m_h$ constraint lie in the narrow range of about $[1.4, 2.6]$, and that the width of the allowed region is essentially constant for $\MSoft \gtrsim 5 \, {\rm TeV}$. The constraint due to the Higgs mass influences the predictions for $\Delta N_{\rm eff}$ only indirectly by narrowing down the possible values of $\MSoft$ and $\tan \beta$. This effect is most visible for $\MSoft \lesssim 2 \, {\rm TeV}$, since such relatively low values of $\MSoft$ require large values of $\tan \beta$ to push the Higgs mass up to around $126 \, {\rm GeV}$. This feature is easy to understand  from~(\ref{eq:mh2}) and explains why the lower  $\Delta N_{\rm eff}$ bound of $1.4$ cannot be saturated for $\MSoft$ below $2 \, {\rm TeV}$. Notice also that the constraint from $m_h$ cuts away the parts of the parameter space with $\Delta N_{\rm eff} \gtrsim 2.6$ and $\MSoft \lesssim 0.5 \, {\rm TeV}$. Both regions are inaccessible because  they correspond to either $\tan \beta \lesssim 2$ or to a too light stop spectrum. We expect that other low-energy constraints (such as~{e.g.}~flavour physics) have an even smaller impact on the limits obtained for $\Delta N_{\rm eff}$ than $m_h$.  The latest Planck measurement of $N_{\rm eff}$~\cite{Ade:2013lta} with (without) the constraint from $H_0$~\cite{Riess:2011yx} gives $\Delta N_{\rm eff} = 0.57 \pm 0.25$ ($\Delta N_{\rm eff} = 0.25 \pm 0.27$). The minimal  value of $\Delta N_{\rm eff} \simeq 1.4$ that is attainable in the MLVS framework thus corresponds to a discrepancy  of about $3.3\sigma$~($4.2\sigma$) between theory and experiment. Deviations in the same ballpark are also found for the $N_{\rm eff}$ extractions by WMAP \cite{Hinshaw:2012aka}, ACT \cite{Hou:2012xq} and SPT \cite{Sievers:2013wk}. These findings basically rule out the MLVS as a model of dark radiation.

\section{Conclusions}
\label{sec:conclusions}

The latest Planck results have ushered in  a new era of precision cosmology. Although these measurements support the standard $\Lambda$CDM  cosmological model, they still leave room for the presence of dark radiation corresponding to up to about  half an extra neutrino species. Other recent experimental determinations of $N_{\rm eff}$ by WMAP, ACT and SPT are within errors all in agreement with the number reported by Planck.

In light of these developments, in this article we have analysed loop corrections to $\Delta N_{\rm eff}$ in the context of sequestered large volume scenarios.  In this class of models, additional contributions to  the effective excess number of neutrinos are an unavoidable consequence of the presence and the interactions of a  light volume modulus $\Phi$:  the decays of this field to the visible sector drive the reheating of the Universe after inflation, while dark radiation  arises from its decays to an ultralight axion partner $a_b$. The only visible-sector decay mode that can compete with the  axion channel is the decay into Higgs pairs induced by  a Giudice-Masiero term. The interplay between the two channels, $\Phi \to a_b \hspace{0.5mm} a_b$ and $\Phi \to H_u H_d$, fixes the relative fraction of dark radiation uniquely in terms of  the coupling strength $Z$ between $\Phi$ and the bilinear $H_u H_d$. Under the assumption that  the  coupling $Z$ is set to 1 at the \compactification scale by means of a shift-symmetric Higgs sector with MSSM matter content, the ratio of branching ratios of visible-sector and hidden-sector decays can then be predicted accurately. At the tree level such a calculation leads to $\Delta N_{\rm eff} \simeq 1.7$, at variance with observation.

Unlike the coupling of the volume modulus to its axion partner, which receives only Planck-suppressed contributions, the $\Phi \hspace{0.5mm} H_u H_d$ coupling is modified by MSSM loops. These radiative corrections induce large logarithms that are formally of ${\cal O} (1)$, and  hence have to be resummed to all orders. In our work, we have calculated the anomalous dimension $\gamma_Z$ of the composite operator $H_u H_d  \hspace{0.5mm} \Box \hspace{0.25mm}  \Phi$ needed to perform such a resummation. We found that the size of the leading-logarithmic corrections to the coupling strength $Z$ depends sensitively on the ratio of the Higgs vacuum expectation values, $\tan \beta$, through the top Yukawa coupling. As a result, loop corrections suppress $\Gamma (\Phi \to H_u H_d)$ for $\tan \beta \lesssim 3$ and $\tan \beta \gtrsim 35$, while the partial decay rate to Higgs pairs is enhanced for all other $\tan \beta$ values. The maximal enhancements occur for $\tan \beta \simeq 10$, but amount to below $10\%$ only.

This simple pattern of suppressions and enhancements is also reproduced by our  high-statistics MSUGRA scan, which includes all relevant two-loop effects. Specifically, we find that in the minimal large volume scenario the values of $\Delta N_{\rm eff}$ that are compatible with a Higgs-boson mass close to $126 \, {\rm GeV}$ all lie in the  range $[1.4,2.6]$. The spread of the predictions is  rather insensitive to the exact values of the MSUGRA parameters $m_0$, $m_{1/2}$ and ${\rm sign} \hspace{0.5mm} \mu$, and is influenced by the Higgs mass requirement  only indirectly because this constraint needs tuning of  $A_0$ and $\tan \beta$. In consequence, it turns out that for moderate values of $\tan \beta$, radiative corrections tend to suppress the tree-level prediction $\Delta N_{\rm eff} \simeq 1.7$. The loop-induced effects are however always small, leading to a robust lower bound  of  $\Delta N_{\rm eff} \gtrsim 1.4$. This limit corresponds to a  $3 \sigma$ to $4 \sigma$ tension between theory and experiment, which essentially excludes the minimal large volume scenario -- MSSM matter content and $Z=1$ -- as a model of dark radiation.

\acknowledgments{JC is funded by the Royal Society with a University Research Fellowship, and by the European Research Council under the Starting Grant ``Supersymmetry Breaking in String Theory". SA and AP are funded by STFC studentships.}

\end{document}